\documentclass[aps,twocolumn,floatfix,amssymb,longbibliography,pra,reprint]{revtex4-2}
\usepackage{graphicx}    
\usepackage{dcolumn}
\usepackage{bm}
\usepackage{booktabs}
\usepackage{caption}
\usepackage{comment}      
\usepackage{graphicx,comment}
\usepackage[english]{babel}
\usepackage{rotating}
\usepackage{braket}
\usepackage{quantikz}
\usepackage{enumerate}
\usepackage{subcaption}
\usepackage{dsfont}
\usepackage{hyperref}

\begin{document}

\title{Time-reversal-based quantum metrology in a central-spin architecture: from SATIN to butterfly echo}

\author{Aditya Kumar}
\email{hrc.adityakumar@gmail.com}
\affiliation{Indian Institute of Science Education \& Research Pune,\\
Dr. Homi Bhabha Road, Pashan, Pune, Maharashtra}

\author{Sayan Choudhury}
\affiliation{Harish Chandra Research Institute, Chhatnag Road,
Jhunsi, Prayagraj, Uttar Pradesh}
\email{sayanchoudhury@hri.res.in}

\begin{abstract}
We investigate quantum-enhanced time-reversal-based metrology using the central spin model, in which a single ancilla spin of magnitude $S_{\rm cen}$ mediates interactions between $N$ bath spins through a star-graph topology. We demonstrate that the SATIN (Signal Amplification Through Time-reversal Interaction) echo protocol can be implemented in this setup to achieve a Fisher information of $N^2/e$ with the collective $\langle J_y\rangle$ readout, corresponding to a $4.34\,\mathrm{dB}$ offset from the Heisenberg limit.  Furthermore, we establish that this platform provides a pathway to perform the butterfly echo protocol by conjugating each evolution step with a random collective bath rotation, thereby achieving axis-agnostic Heisenberg-scaled sensing. Crucially, with a small $N-$independent number of random steps, the butterfly echo reaches a Fisher information of $N^2/4$ ($6\,\mathrm{dB}$ below the Heisenberg limit) for any unknown encoding axis, with the plateau gain tracking the Haar limit. Finally, we demonstrate that the central-spin magnitude $S_{\rm cen}$ provides a powerful knob to reduce the optimal protocol timescale without altering the metrological ceiling or the circuit depth required to reach isotropy. The proposed scheme is relevant to cavity QED and NMR quantum processors, where the central spin architecture arises naturally.
\end{abstract}

\maketitle

%%=============================================================
\section{Introduction}
\label{section:introduction}
%%=============================================================

In classical parameter estimation, $N$ independent sensing particles (such as atoms or photons) measure a phase shift $\phi$ with a statistical uncertainty fundamentally limited by the shot-noise limit, or Standard Quantum Limit (SQL), for which the phase variance scales as $(\Delta\phi)^2 \geq 1/N$~\cite{PhysRevLett.132.190001,MONTENEGRO20251, ghosh2026journeyquantummetrologysensing, agarwal2025quantum,RevModPhys.89.035002, TAYLOR20161, PhysRevResearch.7.013016, https://doi.org/10.1002/qute.202100049}. Quantum resources such as coherence and entanglement provide a route to surpass SQL, and achieve the precision set by the Heisenberg Limit (HL), where $(\Delta\phi)^2 \geq 1/N^2$~\cite{Giovannetti2011, PhysRevLett.96.010401, 10.1063/5.0204102}. However, the highly entangled states required to reach this limit, such as the GHZ and N00N states, are extremely fragile and rapidly degrade in the presence of decoherence and experimental imperfections~\cite{MONTENEGRO20251,PhysRevResearch.4.L012014, doi:10.1126/sciadv.adr1439,Mitchell2004}.

A more robust approach to quantum metrology is provided by spin-squeezed states~\cite{Maccone2020squeezingmetrology}, whose metrological usefulness is quantified by the Wineland parameter $\xi^2 < 1$~\cite{PhysRevA.47.5138,PhysRevA.50.67,PhysRevA.46.R6797}. These states redistribute quantum noise by reducing the variance in one transverse quadrature below the projection-noise limit $N/4$ at the expense of an increased variance in the orthogonal quadrature. Spin squeezing is substantially more resilient to particle loss than
maximally entangled states and can be generated through one-axis
twisting (OAT)~\cite{PhysRevA.47.5138}, cavity-mediated
atom-light interactions~\cite{PhysRevA.89.043837,Colombo_2022},
and Floquet-engineered nonlinear spin dynamics~\cite{32n5-mhk1}.

Despite its robustness, spin squeezing suffers from a fundamental readout problem. Beyond the optimal squeezing time, the state continues to shear around the Bloch sphere, forming non-Gaussian over-squeezed states. These states can retain a large quantum Fisher information (QFI), approaching $F_Q=N^2/2$ in the over-squeezed regime~\cite{PhysRevLett.116.090801}.
This corresponds to a $3.01\,\mathrm{dB}$ offset from the Heisenberg limit. However, extracting this information with restricted collective measurements can be difficult because of the limited resolution of conventional readout schemes~\cite{PhysRevLett.116.090801}.

Time-reversal and echo-based protocols provide a route to extracting metrological information from highly entangled states without single-particle detection~\cite{PhysRevLett.98.223601,PhysRevLett.116.053601,
Colombo_2022,PhysRevA.94.010102,PhysRevLett.117.013001}. In this context, the celebrated SATIN (Signal Amplification Through Time-reversal Interaction) echo protocol has emerged as a powerful route for quantum-enhanced metrology~\cite{PhysRevLett.116.053601,Colombo_2022}. This protocol reverses the Hamiltonian after phase encoding, thereby mapping the accumulated phase information back onto a macroscopic population difference that can be measured with coarse-grained detection~\cite{Colombo_2022}. For the collective $\langle J_y\rangle$ readout used here, the corresponding asymptotic ceiling is $F_Q=N^2/e$, or $4.34\,\mathrm{dB}$ below the Heisenberg limit~\cite{PhysRevLett.116.053601}. This approach has recently been extended to axis-agnostic sensing through the ``butterfly echo'' protocol, which uses approximately Haar-random probe states in the symmetric subspace~\cite{bringewatt2026butterflyechoprotocolaxisagnostic}. We note that this protocol is distinct from the ``butterfly metrology'' protocol~\cite{kobrin2024universalprotocolquantumenhancedsensing,wb5z-y5y5} or other more general chaos-based metrology protocols~\cite{doi:10.1126/science.adg9500,Fiderer2018,asthana2025projected,sharma2026quantum}.

A central experimental challenge is generating the required entanglement within the constraints of realistic experimental platforms. In many settings, including cavity QED and NMR quantum processors, interactions are mediated by a central element such as a cavity photon or a central spin. Motivated by these architectures, we examine time-reversed metrology in the \emph{central spin model} (CSM)~\cite{Nepomechie_2018, PhysRevA.109.052618}. This model can be harnessed for generating entangled states for quantum-enhanced metrology~\cite{Dooley2013collapse,y4g4-bksr,vhx7-gbpt,biswas2025discretetimecrystalsspins}. We demonstrate that both the SATIN and butterfly echo protocols can exhibit Heisenberg scaling of the QFI within the same central-spin Hamiltonian framework. Furthermore, we establish that in the dispersive regime, SATIN outperforms the butterfly echo protocol, when the encoding axis is known. However, the butterfly echo is more robust and it can provide higher metrological gain far away from this regime. Finally, we demonstrate that the magnitude of the central spin can serve as a powerful knob to control the optimal time for the protocol, while keeping the metrological gain ceiling fixed. 

The remainder of this paper is organized as follows. Section~\ref{section:model} introduces the CSM, derives the effective OAT Hamiltonian, and presents both metrological protocols. Section~\ref{section:results} discusses the numerical results for gain dynamics, parameter scaling, and robustness. Finally, Section~\ref{section:conclusion} summarizes our findings. 

%%=============================================================
\section{Model and Protocol}
\label{section:model}
%%=============================================================

\subsection{Central Spin Model}
\label{subsec:CSM}
\begin{figure}
\includegraphics[width=0.5\columnwidth]{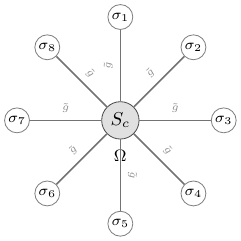}
\caption{%
    The star-graph topology of the central spin model: a single ancilla spin $S_c$ with energy gap $\Omega$ couples uniformly to $N$ bath spins $\sigma_j$ via XX interactions with strength $\tilde{g}$, requiring only $N$ physical bonds with no direct bath-bath coupling.
}
\label{fig:csm_intro}
\end{figure}

The central spin model consists of a single mediator spin of magnitude
$S_{\rm cen}$ coupled uniformly to a bath of $N$ identical spin-$1/2$
sensing spins~\cite{Nepomechie_2018,PhysRevB.99.174308}. The most
general (anisotropic) Hamiltonian is
\begin{equation}
\label{eq:H_XYZ}
    H = \Omega S_z^{(c)}
      + \sum_{j=1}^{N} \!\left(
          A_x S_x^{(c)} \sigma_x^{(j)}
        + A_y S_y^{(c)} \sigma_y^{(j)}
        + A_z S_z^{(c)} \sigma_z^{(j)}
        \right),
\end{equation}
where $S_\alpha^{(c)}$ are the angular momentum operators of the central
spin, $\sigma_\alpha^{(j)}$ are the Pauli operators of the $j$th bath
spin, and $\Omega$ is the central-spin energy gap (detuning). 

We focus on the transverse-isotropic XX case,
$A_x=A_y=\tilde{g}$, with vanishing longitudinal coupling
$A_z=0$. 
Using
$S_x\sigma_x+S_y\sigma_y=\frac{1}{2}(S_+\sigma_-+S_-\sigma_+)$ together with the collective bath operators
$I_\pm^{(b)}=\sum_j\sigma_\pm^{(j)}$, the Hamiltonian reduces to the XX  form~\cite{Nepomechie_2018}:
\begin{equation}
\label{eq:H_XX}
    H = \Omega S_z^{(c)}
      + \frac{\tilde{g}}{2}
        \bigl(S_+^{(c)}I_-^{(b)} + S_-^{(c)}I_+^{(b)}\bigr).
\end{equation}
We take the microscopic single-bond coupling $\tilde{g}$ to be
independent of $N$, corresponding to a fixed microscopic coupling strength as the bath size increases, appropriate for metrological cavity-QED implementations~\cite{Colombo_2022}. We note that ancilla-assisted time-reversal metrology has also been
studied in a related central-spin setting using a longitudinal
$J_z\sigma_z$ probe-ancilla interaction, for which Heisenberg scaling $F_Q=N^2$ can be obtained even from a product probe
~\cite{PhysRevA.110.062425}.Our XX exchange-coupled model, by contrast, transfers excitations between the ancilla and the bath, forbidding that mechanism. Consequently, the metrological gain ceilings we report are specific to the exchange-coupled architecture. 

The homogeneous coupling conserves the total bath spin,
$[H,\mathbf{I}^2]=0$, confining the dynamics to the symmetric
Dicke subspace of dimension $N+1$~\cite{Nepomechie_2018}. Consequently,
the effective Hilbert-space dimension is reduced from $2^N$ to
$(2S_{\rm cen}+1)(N+1)$, enabling exact simulations for systems
containing up to $N=1000$ bath spins.

\begin{figure*}[htbp]
\centering
\includegraphics[width=1.9\columnwidth]{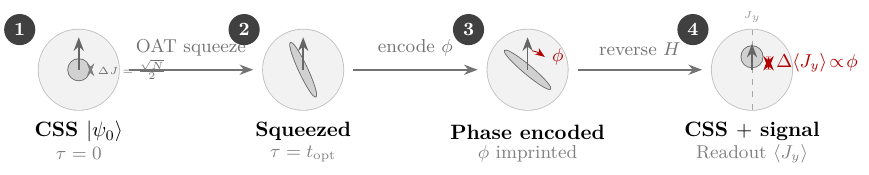}
   \caption{%
    SATIN echo protocol implemented using the full central-spin
    Hamiltonian. \textbf{(1)}~The bath is initialized in a coherent
    spin state (CSS), with the central spin in its lowest-energy
    state. \textbf{(2)}~Forward evolution under the full Hamiltonian
    generates the entangled probe state; in the dispersive regime,
    this evolution is described approximately by the effective OAT
    Hamiltonian $H_{\rm eff}^{(b)}$ (Eq.~\eqref{eq:Heff_OAT}).
    \textbf{(3)}~The unknown phase $\phi$ is encoded.
    \textbf{(4)}~Time reversal under the full Hamiltonian returns
    the bath to its initial CSS at $\phi=0$, while a nonzero phase
    produces a displacement of $\langle J_y\rangle$ proportional
    to $\phi$. The effective-OAT prediction for the signal is given
    by Eq.~\eqref{eq:signal_OAT}.
}
\label{fig:echo_protocol}
\end{figure*}

In the dispersive regime,
$\Omega\gg\tilde{g}\sqrt{N}$, the central spin remains only virtually
excited and mediates an effective interaction between the bath spins. A
second-order Schrieffer--Wolff transformation (derived in
Appendix~\ref{app:eff_ham}) yields the effective Hamiltonian
\begin{equation}
\label{eq:Heff_OAT}
    H_{\rm eff}^{(b)}
    =
    \chi\bigl(J_z\bigr)^2,
\end{equation}
with
\begin{equation}
\label{eq:chi_def}
    \chi
    =
    \frac{\tilde{g}^2S_{\rm cen}}{2\Omega}.
\end{equation}
The effective Hamiltonian is identical in form to the one-axis twisting (OAT) Hamiltonian, $H_{\rm OAT}=\chi J_z^2$~\cite{PhysRevA.47.5138}.
In this work, both the SATIN and butterfly echo protocols are implemented using the full central-spin Hamiltonian $H$ in Eq.~\eqref{eq:H_XX}. In the dispersive regime, $\Omega\gg\tilde{g}\sqrt{N}$, the full CSM dynamics reduce approximately to the corresponding OAT dynamics generated
by $H_{\rm eff}^{(b)}$. Since $\chi\propto S_{\rm cen}$, increasing the central-spin magnitude accelerates the effective squeezing dynamics without changing the ultimate metrological ceiling. For $\tilde{g}=2$, Eq.~\eqref{eq:chi_def} reduces to $\chi=2S_{\rm cen}/\Omega$. The argument outlined so far suggests that the CSM architecture would enable time-reversed metrology protocols in the dispersive regime. We emphasize, however, that in this work, we demonstrate that the CSM can be harnessed for time-reversed metrology, even away from this regime, thereby demonstrating the versatility of this architecture.\\

Since time reversal is at the core of our protocol, it is instructive
to describe how it is implemented in the CSM. In the ideal coherent
model considered here, the inverse unitary $U^\dagger$ is implemented
by a second evolution under the full Hamiltonian $H$ with the sign
reversed, so that $U^\dagger=e^{+iH\tau}=e^{-i(-H)\tau}$.
For the XX central-spin Hamiltonian, this sign reversal is achieved
by applying a $\pi$ pulse about the $y$ axis to the central spin and
a collective $\pi$ pulse about the $x$ axis to the bath:
\[
R=e^{-i\pi S_y^{(c)}}e^{-i\pi J_x}.
\]
The two transverse $\pi$ rotations give
$R S_z^{(c)}R^\dagger=-S_z^{(c)}$ and reverse the sign of the
exchange term, so that
\[
RHR^\dagger=-H,
\qquad
R e^{-iH\tau}R^\dagger=e^{+iH\tau}=U^\dagger.
\]
This conjugation-based construction follows the general strategy of
realizing reverse evolution by transforming the Hamiltonian to its
negative~\cite{Braumüller2022,wb5z-y5y5}. Hamiltonian time reversal
has also been implemented experimentally in dipolar interacting
many-body spin systems, demonstrating the feasibility of reversing
complex many-body dynamics through controlled changes of the
interaction Hamiltonian~\cite{PhysRevResearch.6.033197}. This recipe
requires $A_z=0$, since a longitudinal
$A_z S_z^{(c)}\sigma_z^{(j)}$ coupling is invariant under both flips
and would survive unchanged; this is already our working assumption.
Crucially, the reversal is of the \emph{full} Hamiltonian $H$, not
merely the effective OAT approximation, so the cancellation
$U^\dagger(\phi=0)U=\mathds{1}$ holds exactly at all orders in the
Schrieffer--Wolff expansion. No dispersive approximation is required
for the echo to close.
The near-Haar plateau observed at $\Omega=100$ is instead a separate numerical result discussed in Sec.~\ref{subsec:results_butterfly}.

\subsection{Standard Echo Protocol (SATIN)}
\label{subsec:standard_echo}

Throughout this subsection, we use the standard OAT notation~\cite{PhysRevA.47.5138}, writing
$J_\alpha \equiv I_\alpha^{(b)}$ and $J=N/2$ for the collective bath spin operators. The superscript $(b)$ is reintroduced where needed for clarity.

To access the large QFI stored in over-squeezed states, we employ the
SATIN echo protocol (Fig.~\ref{fig:echo_protocol})
~\cite{PhysRevLett.116.053601,Colombo_2022}. The protocol is implemented
using the full central-spin Hamiltonian $H$ in Eq.~\eqref{eq:H_XX}.

\begin{enumerate}
    \item \textbf{Entanglement generation:} Prepare the central spin in
          its lowest-energy state and the bath in the coherent spin state
          $|{+}x\rangle^{\otimes N}$. Evolve the joint system under the
          full Hamiltonian for a time $t$,
          \[
          |\Psi(t)\rangle
          =
          e^{-iHt}
          \bigl(|-S_{\rm cen}\rangle_c
          \otimes |{+}x\rangle_b^{\otimes N}\bigr).
          \]

    \item \textbf{Phase encoding:} Apply the unknown rotation
          \[
          |\Psi(\phi,t)\rangle
          =
          \bigl(\mathbb{I}_c\otimes e^{-i\phi J_y}\bigr)
          |\Psi(t)\rangle.
          \]

    \item \textbf{Time-reversal and readout:} Apply the inverse evolution
          $e^{+iHt}$ using the physical Hamiltonian-inversion procedure
          described above, and measure $\langle J_y\rangle$.
\end{enumerate}

At $\phi=0$, the forward and reverse evolutions cancel exactly, returning the joint system to its initial state and the bath to its initial coherent spin state.
In the dispersive regime, the effective OAT description gives the
analytical prediction for the signal slope:

\begin{equation}
\label{eq:signal_OAT}
    \mathcal{S}
    =
    \frac{N}{2}(N-1)
    \sin\!\left(\frac{2S_{\rm cen}t}{\Omega}\right)
    \cos^{N-2}\!\left(\frac{2S_{\rm cen}t}{\Omega}\right),
\end{equation}
where we have used $\chi_{\rm eff}=2S_{\rm cen}/\Omega$, corresponding to $\tilde{g}=2$. The phase sensitivity obtained from this analytical OAT description is given by
\begin{equation}
\label{eq:delta_phi}
    \Delta\phi
    =
    \frac{\sqrt{N/4}}
         {\mathcal{S}(t)},
\end{equation}
giving the metrological gain relative to the SQL,
\begin{equation}
\label{eq:gain_dB}
    \mathcal{G}
    =
    10\log_{10}\!\left(
    \frac{1}{N(\Delta\phi)^2}
    \right)
    \quad (\mathrm{dB}).
\end{equation}

At the optimal evolution time,
$t_{\rm opt}=\Omega/(2S_{\rm cen}\sqrt{N})$
(Appendix~\ref{app:topt}), the gain reaches to a maximum Fisher information associated with the specified collective readout of
$10\log_{10}(N)-4.34\,\mathrm{dB}$. 

\subsection{Butterfly Echo Protocol}
\label{subsec:butterfly}

The SATIN protocol prepares a highly anisotropic probe state, with maximum sensitivity to rotations about a known axis. If the encoding axis $\hat{n}$ is unknown, its performance degrades substantially. To overcome this limitation, we employ the recently proposed butterfly echo protocol~\cite{bringewatt2026butterflyechoprotocolaxisagnostic}, which enables axis-agnostic Heisenberg-scaled sensing. Related axis-agnostic probe states have been studied in the context of quantum rotosensing~\cite{PhysRevA.95.052125} and random bosonic states~\cite{PhysRevX.6.041044}; the butterfly echo provides an experimentally accessible route to generating such states.

\begin{figure*}[htbp]
\centering
\includegraphics[width=2\columnwidth]{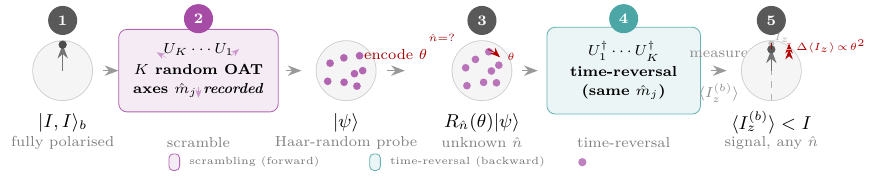}
   \caption{%
    Butterfly echo protocol implemented using the central-spin model.
    \textbf{(1)}~The central spin is initialized in its lowest-energy
    state and the bath in the fully polarized Dicke state $|I,I\rangle_b$.
    \textbf{(2)}~$K=8$ random bath rotations are used to conjugate
    successive evolutions under the full central-spin Hamiltonian.
    In the dispersive regime, these steps approximate the random-axis
    OAT circuit and scramble the probe into an approximately Haar-random
    state. \textbf{(3)}~An unknown rotation $R_{\hat{n}}(\theta)$ about
    an arbitrary axis $\hat{n}$ encodes the signal.
    \textbf{(4)}~The inverse sequence of full CSM unitaries is applied
    using the same recorded random axes.
    \textbf{(5)}~The bath returns to full polarization at zero encoded
    phase, while a nonzero phase produces a measurable change in
    $\langle J_z\rangle$.
}
\label{fig:butterfly_protocol}
\end{figure*}

\subsubsection{The Axis-Agnosticity Problem}

Consider encoding a rotation
$R_{\hat{n}}(\theta)=e^{-i\theta J_{\hat{n}}}$
about an \emph{unknown} axis $\hat{n}$. In the single-shot, non-adaptive regime, the encoding is described by the mixed-unitary (depolarizing) channel~\cite{bringewatt2026butterflyechoprotocolaxisagnostic}
\begin{equation}
\label{eq:depolarising_channel}
    \rho_\theta
    =
    \mathcal{M}_\theta[\rho]
    =
    \frac{1}{4\pi}
    \int\!\mathrm{d}\hat{n}\;
    R_{\hat{n}}(\theta)\,
    \rho\,
    R_{\hat{n}}^\dagger(\theta),
\end{equation}
which, to second order in $\theta$, yields the QFI~\cite{bringewatt2026butterflyechoprotocolaxisagnostic}
\begin{equation}
\label{eq:QFI_depolarising}
    F(\theta)
    =
    \frac{4}{3}
    \sum_{j\in\{x,y,z\}}
    \mathrm{Var}(J_j).
\end{equation}
Heisenberg-limited axis-agnostic sensing therefore requires maximizing the sum of spin variances along all three directions simultaneously. States with this property, known as anticoherent states, have been studied extensively~\cite{10.21468/SciPostPhysCore.9.1.001}.

For a Haar-random probe state in the symmetric Dicke subspace, the
ensemble-averaged QFI of the unknown-axis channel is
\begin{equation}
\label{eq:QFI_haar}
    \overline{F_{\rm Haar}}
    =
    \frac{N(N+2)}{3}
    \approx
    \frac{N^2}{3}.
\end{equation}
This gives the asymptotic quantum Cramér--Rao scaling
$\Delta\theta^{-1}\sim N/\sqrt{3}$, whereas the butterfly echo achieves
$\Delta\theta^{-1}\sim N/2$ with its collective readout
~\cite{bringewatt2026butterflyechoprotocolaxisagnostic,PhysRevX.6.041044,Collins2021TheWC}.
The reduction from the channel QFI to the echo-extracted QFI is a
separate property of the collective echo readout, which extracts a
fraction $3/4$ of the channel QFI:
\[
F_{\rm echo}=\frac{3}{4}F_{\rm channel}.
\]
The resulting Haar-limit metrological gain is
\begin{equation}
\label{eq:haar_limit}
    \mathcal{G}_{\rm Haar}^{\rm echo}
    =
    10\log_{10}\!\left(\frac{N}{4}\right).
\end{equation}
The butterfly echo therefore reaches the Heisenberg exponent while remaining exactly
$6\,\mathrm{dB}$ below the Heisenberg limit. The corresponding gain differs from the SATIN protocol only by a constant prefactor ($c=e$ versus $c=4$), so both protocols exhibit the same asymptotic scaling,
$F_Q\propto N^2$.

\subsubsection{Probe Preparation and Convergence}

The butterfly echo requires a probe state with approximately isotropic
spin fluctuations in the symmetric Dicke subspace. Although exact
Haar-random state preparation is generally costly, previous work has
shown that a low-depth scrambling sequence can reach near-Haar
metrological performance with a number of steps that is approximately
independent of the system size~\cite{bringewatt2026butterflyechoprotocolaxisagnostic}.
In our implementation, the probe is prepared by applying successive
randomly rotated evolution steps under the full central-spin
Hamiltonian, as described below. Our numerical results show that
$K=8$ steps suffice to reach the Haar-limit performance within the
numerical accuracy reported.

\subsubsection{CSM Implementation via Random Bath Rotations}

The central spin model generates OAT only about the fixed
$\hat{z}$ axis. To realize random-axis OAT, each CSM evolution step is conjugated by random collective rotation of the bath:
\begin{equation}
\label{eq:butterfly_CS_step}
    U_j(\hat{m}_j)
    =
    \bigl[
    \mathds{1}_c
    \otimes
    R_{\rm bath}(\hat{m}_j)
    \bigr]
    e^{-iH\tau}
    \bigl[
    \mathds{1}_c
    \otimes
    R_{\rm bath}^\dagger(\hat{m}_j)
    \bigr],
\end{equation}
where
\[
R_{\rm bath}(\hat{m}_j)
=
\exp(i\alpha_j\hat{m}_j\cdot\mathbf{I}^{(b)})
\]
is a collective bath rotation with
$\hat{m}_j$ uniformly distributed over
$\mathbb{S}^2$ and
$\alpha_j\in[0,2\pi)$.
This conjugated evolution under the full central-spin Hamiltonian
is the physical scrambling step used in our simulations. In the
dispersive regime, $\Omega\gg\tilde{g}\sqrt{N}$, it reduces
approximately to the usual random-axis OAT evolution,
\begin{equation}
\label{eq:butterfly_effective}
    U_j(\hat{m}_j)
    \approx
    \exp\!\left[
    -i\chi_{\rm eff}\tau(J_{\hat{m}_j})^2
    \right].
\end{equation}
This effective description provides a connection to the standard
butterfly protocol, while the full CSM evolution also allows us to
investigate regimes beyond the dispersive approximation. The random
axes are recorded classically, allowing the inverse sequence
$U^\dagger=U_1^\dagger\cdots U_K^\dagger$ to be applied during
readout. The effects of deviations from the effective description
are analyzed in App.~\ref{app:butterfly} and Tab.~\ref{tab:epsilon}.

\subsubsection{Full Protocol}

The butterfly echo protocol proceeds as follows.

\begin{enumerate}
    \item \textbf{Initialization:} Prepare the bath in the fully polarized Dicke state
    $|\downarrow\rangle_c\otimes|I,I\rangle_b$.

    \item \textbf{Forward scrambling:} Apply $K$ random CSM unitaries
    $U_j(\hat{m}_j)$
    [Eq.~\eqref{eq:butterfly_CS_step}]
    to generate the approximate Haar-random probe state
    $|\psi\rangle
    =
    U_K\cdots U_1|I,I\rangle_b$.

    \item \textbf{Phase encoding:} Apply the unknown rotation
    $e^{-i\theta J_{\hat n}}|\psi\rangle$.

    \item \textbf{Reverse scrambling:} Apply
    $U_1^\dagger\cdots U_K^\dagger$
    and measure
    $\langle J_z\rangle$.
\end{enumerate}

For $\theta=0$, the forward and reverse circuits cancel exactly, restoring the fully polarized Dicke state with
$\langle J_z\rangle=I=N/2$.
For finite $\theta$, the signal follows~\cite{bringewatt2026butterflyechoprotocolaxisagnostic}
\begin{equation}
\label{eq:butterfly_signal}
    \frac{\overline{\langle J_z\rangle}}{I}
    \xrightarrow{N\rightarrow\infty}
    \mathrm{sinc}^2(I\theta),
\end{equation}
where the overbar denotes averaging over random circuit realizations.

The metrological gain is extracted numerically as
\begin{equation}
\label{eq:butterfly_gain}
    \mathcal{G}
    =
    10\log_{10}\!\left[
    \frac{1}{N}
    \frac{\left[2(I-\langle J_z\rangle)/\theta_{\rm b}\right]^2}
         {\mathrm{Var}(J_z)}
    \right],
\end{equation}
using a bias angle $\theta_{\rm b}=10^{-4}$ as a stated operating convention. Because the butterfly signal is quadratic near $\theta=0$, the extracted sensitivity depends on this choice; we have verified that the plateau gain extracted with this convention reproduces the analytic Haar limit of Eq.~\eqref{eq:haar_limit} within the numerical precision reported.
Each data point is averaged over
$n_{\rm circ}=50$
independent random circuit realizations.

We note that unlike the SATIN protocol, whose output signal
$\langle J_y\rangle\propto\phi$
is linear in the encoded phase, the butterfly echo produces a quadratic response near
$\theta=0$ because the fully polarized Dicke state is an eigenstate of $I_z$. The sensitivity is therefore determined by the curvature of the $\mathrm{sinc}^2(I\theta)$ signal rather than its slope.

%%=============================================================
\section{Results and Discussion}
\label{section:results}
%%=============================================================

We now investigate whether the star-graph topology of the CSM can sustain Heisenberg-scaled sensing, and how the control parameters $N$, $\Omega$, and $S_{\rm cen}$ govern the metrological gain and timescale of both the SATIN and butterfly echo protocols. All CSM numerical results reported in this section are obtained by exact diagonalization of the full central-spin Hamiltonian $H$ in Eq.~\eqref{eq:H_XX} within the symmetric Dicke subspace of dimension $(2S_{\rm cen}+1)(N+1)$. The effective OAT model is used to derive analytical predictions and as a benchmark for interpreting the results in the dispersive regime. Unless stated otherwise, we use $\tilde{g}=2$, $S_{\rm cen}=0.5$, $N\in\{100,200,300,400,500,800,1000\}$, and
$\Omega\in\{100,300,500,1000,2000\}$. Results for $S_{\rm cen}\in\{1.0,1.5\}$ are presented in Apps.~\ref{app:scen} and~\ref{app:butterfly}.

\subsection{Standard Echo: Gain Dynamics and Parameter Scaling}
\label{subsec:results_echo}

\begin{figure*}[htbp]
    \centering
    \includegraphics[width=0.70\textwidth]{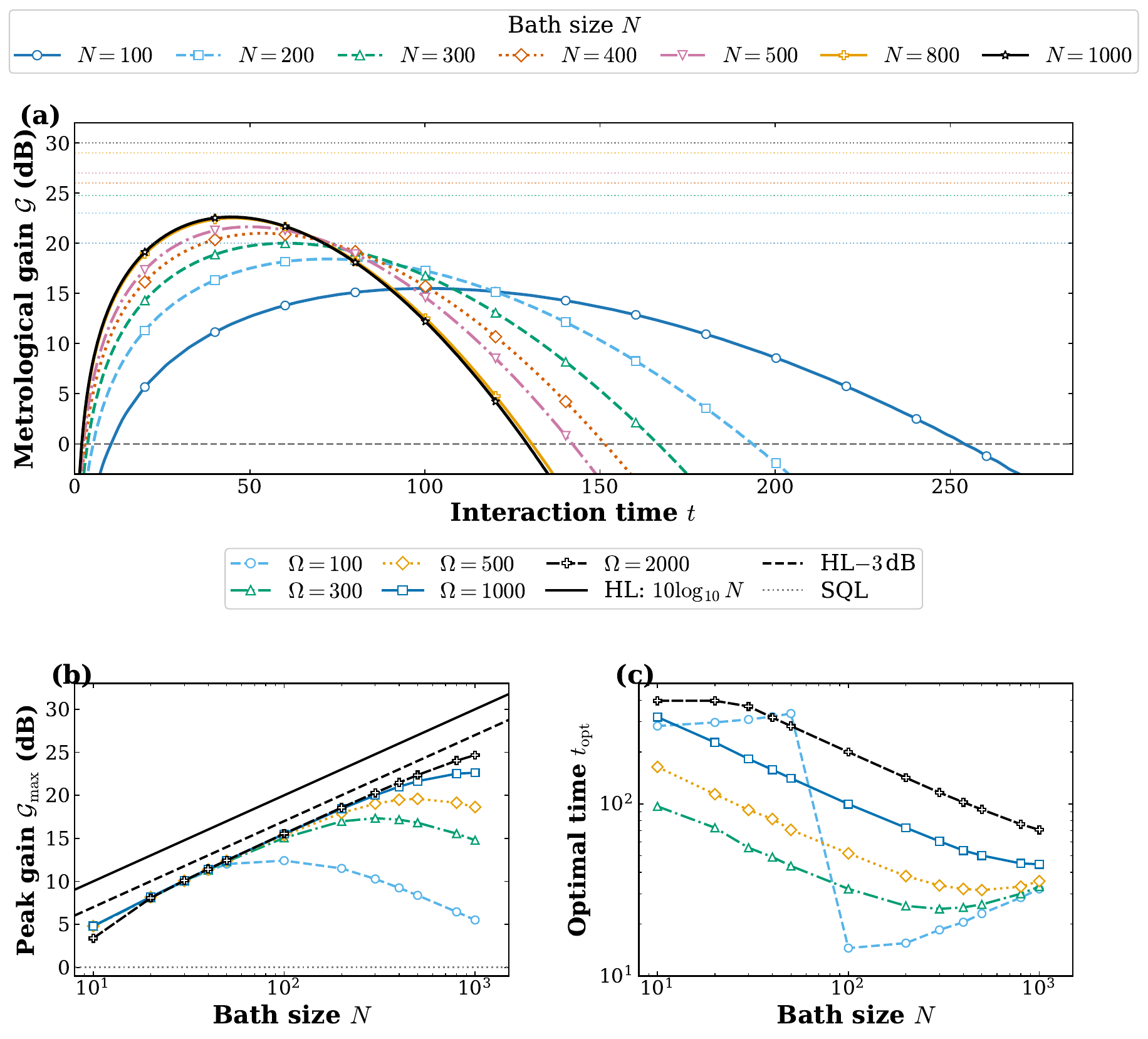}
    \caption{%
        Standard echo (SATIN) protocol via the central spin model,
        $S_{\rm cen}=0.5$, $\tilde{g}=2$.
        \textbf{(a)}~Metrological gain $\mathcal{G}$ vs time $t$ for
        $\Omega=1000$ and $N\in\{100,200,300,400,500,800,1000\}$.
        Bell-shaped curves show the characteristic SATIN gain profile;
        the dashed lines indicate the squeezed-state parity-readout benchmark $\mathrm{HL}-3\,\mathrm{dB}$, while the ideal SATIN collective-readout ceiling is $\mathrm{HL}-4.34\,\mathrm{dB}$; the residual deficit relative to the ideal SATIN ceiling and its growth at large $N$ are discussed in Sec.~\ref{subsec:results_echo}.
        Coloured dotted lines: individual Heisenberg limits $10\log_{10}N$.
        Grey dashed: SQL ($0\,\mathrm{dB}$).
        \textbf{(b)}~Peak gain vs $N$ for
        $\Omega\in\{100,300,500,1000,2000\}$.
        The fitted slopes range from $7.2\,\mathrm{dB/decade}$ at $\Omega=300$ to $9.2\,\mathrm{dB/decade}$ at $\Omega=2000$, approaching the Heisenberg slope of $10\,\mathrm{dB/decade}$. For $\Omega=100$, the turnover at large $N$ marks the breakdown of the dispersive approximation.
        \textbf{(c)}~Optimal time $t_{\rm opt}$ vs $N$ (log-log).
        Dotted lines: power-law fits; fitted exponent
        ${\propto}N^{-0.46}$ for $\Omega=1000$ is consistent with the
        analytical prediction Eq.~\eqref{eq:topt_CSM}.
    }
    \label{fig:echo_dynamics}
\end{figure*}

\paragraph{Gain profile and the role of $N$.} Figure~\ref{fig:echo_dynamics}(a) shows the metrological gain as a function of interaction time for seven bath sizes at fixed $\Omega=1000$. Each curve exhibits the characteristic SATIN profile: the gain initially increases as OAT generates metrological entanglement, reaches a maximum at $t=t_{\rm opt}$, and subsequently decreases as continued twisting rotates the squeezed quadrature away from the optimal measurement axis, reducing the projection of the encoded phase onto the measured observable. Each curve approaches the ideal squeezed-state benchmark
$10\log_{10}(N)-3\,\mathrm{dB}$ (dashed lines), corresponding
to the maximum gain attainable with parity readout. This
benchmark is distinct from the SATIN echo limit, since SATIN
uses a collective $\langle J_y\rangle$ readout rather than
parity detection. For the SATIN protocol, the corresponding
asymptotic Fisher information associated with the specified collective readout is $N^2/e$, or
$10\log_{10}(N)-4.34\,\mathrm{dB}$ below the Heisenberg limit. The remaining deficit has two distinct origins. The first is the intrinsic $1.34\,\mathrm{dB}$ readout penalty
associated with the collective $\langle J_y\rangle$ measurement,
which reduces the Fisher information accessible with the collective
readout from the squeezed-state parity-readout benchmark $N^2/2$
to the SATIN ceiling $N^2/e$ (Appendix~\ref{app:deltaphi})
~\cite{PhysRevLett.116.090801}. The second arises from finite Schrieffer--Wolff (SW) corrections in the CSM. Because the time-reversal is of the full Hamiltonian (not just the OAT approximation), the bath returns \emph{exactly} to a coherent spin state at $\phi=0$ and the readout noise is precisely the projection noise $\sqrt{N/4}$. The SW deficit instead resides in the \emph{signal slope}: with the full CSM propagator rather than ideal OAT, part of the accumulated correlation is stored in the mediator spin during the forward evolution and never converts into $\langle J_y\rangle$ displacement, reducing the slope $\mathcal{S}$ relative to the ideal value. This mechanism has the same $\Omega$ dependence,
$\epsilon_{\rm SW}\sim4(\tilde g\sqrt{N}/\Omega)^2$
[Table~\ref{tab:epsilon}], and explains why the butterfly echo-which needs only isotropy within the Dicke subspace-is insensitive to the same SW errors that degrade the SATIN signal. Increasing $\Omega$ from $1000$ to $2000$ reduces the additional deficit at $N=1000$ from $2.42\,\mathrm{dB}$ to $0.77\,\mathrm{dB}$, consistent with the expected $1/\Omega^2$ scaling. Conversely, for small $N$, where $\tilde g\sqrt N\ll\Omega$,
the CSM results approach the ideal SATIN echo limit, confirming that the residual discrepancy originates from finite-order SW corrections rather than the protocol itself. \paragraph{Effect of $\Omega$ on peak gain and timing.} Figure~\ref{fig:echo_dynamics}(b) shows the peak gain as a function of $N$ for five detunings. In the dispersive regime ($\Omega\ge300$), the curves are nearly coincident, with fitted slopes ranging from $7.2\,\mathrm{dB/decade}$ at $\Omega=300$ to $9.2\,\mathrm{dB/decade}$ at $\Omega=2000$, approaching the Heisenberg value of $10\,\mathrm{dB/decade}$. For $\Omega=100$, where the dispersive approximation breaks down for $N\gtrsim100$, the peak gain in Fig.~\ref{fig:echo_dynamics}(b) turns over and \emph{falls} with increasing $N$: the Heisenberg-like advantage collapses as the bath size grows and the central spin begins exchanging real angular momentum with the bath rather than acting as a virtual mediator. This near-resonant regime is therefore not a viable operating point for Heisenberg-scaled sensing, and serves primarily to illustrate the lower boundary of the dispersive regime. The detuning also controls the protocol timescale. From Eq.~\eqref{eq:topt_CSM}, $t_{\rm opt}\propto\Omega/(2S_{\rm cen}\sqrt N)$, so increasing $\Omega$ shifts the gain profile to later times without significantly changing its peak value. This behaviour is confirmed in Fig.~\ref{fig:echo_dynamics}(c), where the fitted scaling exponent is $\alpha\approx0.46$ for $\Omega=1000$, close to the OAT prediction $\alpha=0.5$ (Appendix~\ref{app:topt}), with the small deviation attributable to finite-order SW corrections. Thus, $\Omega$ primarily acts as a timing parameter while leaving the asymptotic metrological performance unchanged. \paragraph{Effect of $S_{\rm cen}$.} Results for $S_{\rm cen}\in\{1.0,1.5\}$, presented in Appendix~\ref{app:scen} (Fig.~\ref{fig:standard_echo_comparison}), show that the peak gain is essentially unchanged from the $S_{\rm cen}=0.5$ case, whereas $t_{\rm opt}$ is reduced by factors of two and three, respectively. This follows directly from $\chi_{\rm eff}=\tilde g^2S_{\rm cen}/(2\Omega)$ [Eq.~\eqref{eq:chi_def}]: increasing the central-spin magnitude accelerates the OAT dynamics without changing the optimal squeezing angle or the maximum achievable QFI. This decoupling provides a useful control strategy because the protocol duration can be reduced by increasing the mediator spin without sacrificing metrological performance. Together, these results establish a clear hierarchy among the control parameters: $N$ determines the metrological ceiling, $\Omega$ sets the protocol timescale, and $S_{\rm cen}$ controls the squeezing rate, providing three independent control parameters for optimizing the protocol.

\subsection{Butterfly Echo: Gain Dynamics and Parameter Scaling}
\label{subsec:results_butterfly}

\begin{figure*}[htbp]
    \centering
    \includegraphics[width=0.75\textwidth]{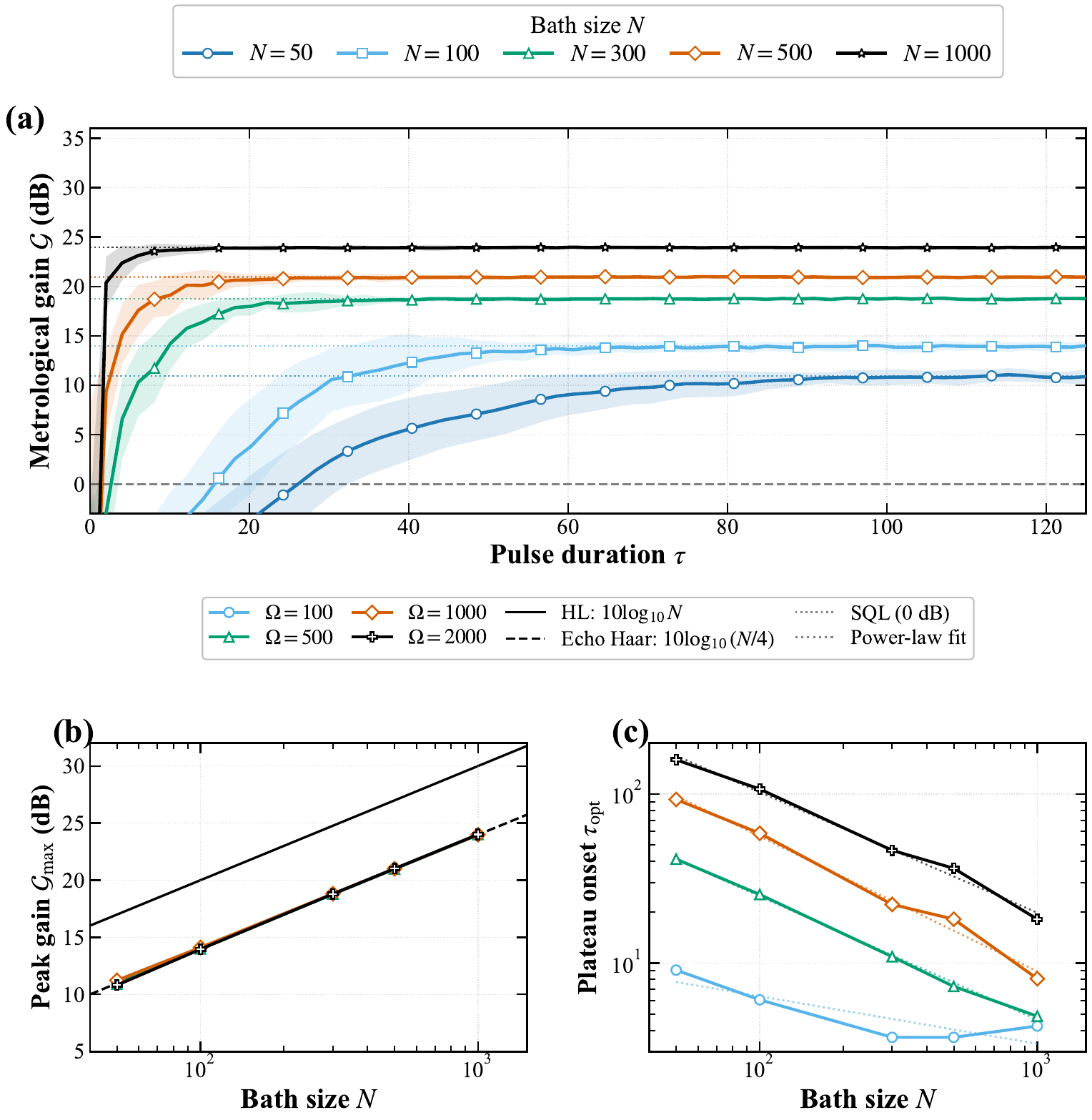}
    \caption{%
        Butterfly echo protocol via the central spin model,
        $S_{\rm cen}=0.5$, $\tilde{g}=2$, $K=8$.
        Shaded bands: $\pm 1\sigma$ over 50 random circuit realisations.
        \textbf{(a)}~Gain $\mathcal{G}$ vs pulse duration $\tau$ for
        $\Omega=1000$ and $N\in\{50,100,300,500,1000\}$.
        Each curve rises to a stable plateau at the echo Haar limit
        $10\log_{10}(N/4)$ (Eq.~\eqref{eq:haar_limit}, coloured dotted lines),
        in contrast to the bell-shaped profile of Fig.~\ref{fig:echo_dynamics}(a).
        Grey dashed: SQL.
        \textbf{(b)}~Peak plateau gain vs $N$ for
        $\Omega\in\{100,500,1000,2000\}$.
        All four $\Omega$ values collapse onto the Haar limit (dashed),
        with fitted slopes $9.8$-$10.1\,\mathrm{dB/decade}$;
        the small residual gap from the Haar limit for $\Omega=100$ at large $N$
        reflects the dispersive approximation error (Table~\ref{tab:epsilon}).
        \textbf{(c)}~Plateau onset $\tau_{\rm opt}$ vs $N$ (log-log),
        consistent with the scaling law.
        Fitted exponent ${\propto}N^{-0.80}$ for $\Omega=1000$;
        dotted lines: power-law fits per $\Omega$.
    }
    \label{fig:butterfly_dynamics}
\end{figure*}

\paragraph{Plateau dynamics.}

Figure~\ref{fig:butterfly_dynamics}(a) shows that the butterfly echo exhibits dynamics qualitatively different from the SATIN protocol. Rather than displaying a bell-shaped gain profile, the metrological gain increases monotonically before saturating at a stable plateau that persists for the remainder of the evolution.

This behaviour reflects the different mechanisms underlying the two protocols. In SATIN, the interaction time determines the squeezing axis, giving a unique optimal time beyond which continued evolution rotates the squeezed quadrature away from the measurement axis. In contrast, each butterfly pulse rotates the OAT axis to a new random direction, progressively scrambling the probe into an isotropic state within the symmetric Dicke subspace. Once this isotropic state is established (typically after $K\approx8$ steps), additional evolution neither increases nor decreases the metrological gain, resulting in the observed plateau. We also note that because the time-reversal is of the full Hamiltonian $H$, the cancellation at zero signal ($\theta=0$) is exact at all orders; the bath returns precisely to the initial state regardless of the Schrieffer--Wolff accuracy, confirming that the plateau is a genuine physical result.

\paragraph{Saturation at the echo Haar limit.}

All five curves in Fig.~\ref{fig:butterfly_dynamics}(a) saturate within
$0.3\,\mathrm{dB}$ of the echo Haar limit,
$10\log_{10}(N/4)$
[Eq.~\eqref{eq:haar_limit}], in agreement with the theoretical prediction. The shaded $\pm1\sigma$ bands, obtained from 50 independent circuit realizations, are broad during the transient regime but collapse rapidly once the plateau is reached. This self-averaging reflects convergence toward an approximate unitary $2$-design, for which the circuit-to-circuit fluctuations satisfy
$\mathrm{std}(F_Q)/\overline{F_Q}\sim N^{-1/2}\rightarrow0$
(Appendix~\ref{app:butterfly}). For
$N\ge100$,
the plateau varies by less than
$0.3\,\mathrm{dB}$
across all realizations, indicating that a single random circuit can be reused as a fixed sensing resource.

\paragraph{Effect of $\Omega$.}

Figure~\ref{fig:butterfly_dynamics}(b) shows that the plateau gain for
$\Omega\in\{100,500,1000,2000\}$
collapses onto a single curve following
$10\log_{10}(N/4)$,
with fitted slopes of
$9.8$-$10.1\,\mathrm{dB/decade}$
for
$\Omega\ge500$.
This contrasts with the SATIN protocol, where the gain ceiling exhibits a measurable dependence on the detuning.

The origin of this behaviour is that the butterfly metrological ceiling is determined primarily by the dimension of the symmetric Dicke subspace rather than by the accuracy of the OAT mapping. The Haar-averaged QFI,
$\overline{F_Q}=N(N+2)/3$,
depends only on the subspace dimension~\cite{bringewatt2026butterflyechoprotocolaxisagnostic}. Consequently, moderate Schrieffer--Wolff errors do not prevent convergence toward an approximately isotropic state. Even for $\Omega=100$, where the dispersive approximation is no longer well satisfied, the numerical results remain within $0.5\,\mathrm{dB}$ (\ref{tab:epsilon})
of the Haar limit for the system sizes studied. This indicates that the
butterfly protocol can retain near-Haar performance in the parameter
range investigated, despite the breakdown of the leading-order OAT
mapping.

The detuning therefore plays a role similar to that in SATIN by setting the protocol timescale rather than the asymptotic metrological performance. Figure~\ref{fig:butterfly_dynamics}(c) shows that
$\tau_{\rm opt}$
increases approximately linearly with
$\Omega$,
consistent with
$\tau_{\rm opt}\propto\Omega/(2S_{\rm cen}\tilde g^2N^\alpha)$.
The fitted exponent,
$\alpha\approx0.80$
for
$\Omega=1000$,
is larger than the SATIN value because each random OAT step must produce a full wrap of the coherent state around the Bloch sphere before the next random rotation is applied.

\paragraph{Effect of $S_{\rm cen}$.}

Appendix~\ref{app:butterfly}
(Fig.~\ref{fig:appendix_butterfly_higher_spins})
shows that increasing
$S_{\rm cen}$
from
$0.5$
to
$1.0$
or
$1.5$
leaves the plateau height unchanged while reducing
$\tau_{\rm opt}$
by factors of two and three, respectively. Likewise, the convergence of the plateau with circuit depth remains unchanged, reaching saturation at
$K\approx8$ for all $S_{\rm cen}$ values studied here
(Appendix~Fig.~\ref{fig:plateau_vs_K}).
Thus, the mediator spin controls only the protocol speed, while the plateau height and the required circuit depth remain unchanged.

\paragraph{Comparison with SATIN.}

The maximum Fisher information obtained with the SATIN and butterfly readouts is $N^2/e$ and $N^2/4$, respectively. The physical origin of their difference is discussed in Sec.~\ref{subsec:gap_hierarchy}: the reduction is a consequence of axis independence rather than of the central-spin implementation or the finite circuit depth.

\subsection{The Gap Hierarchy and Unified View}
\label{subsec:gap_hierarchy}

For the class of interaction-based readout protocols with collective measurement and an entangled probe studied here, the metrological ceilings form the hierarchy:
\begin{equation}
\underbrace{F_Q = N^2}_{\substack{\text{Heisenberg limit}}}
>
\underbrace{\frac{N^2}{2}}_{\substack{\text{OAT state QFI}\\\text{(parity detection)}}}
>
\underbrace{\frac{N^2}{e}}_{\substack{\text{SATIN}\\\langle J_y\rangle\text{ readout}}}
>
\underbrace{\frac{N^2}{4}}_{\substack{\text{butterfly echo}\\\text{(any }\hat{n})}}.
\label{eq:ceiling_hierarchy}
\end{equation}
We note that differently-used ancillas can attain $F_Q=N^2$ under different resource assumptions (resolved readout, strong coupling, known axis, longitudinal coupling)~\cite{PhysRevA.110.062425}; the hierarchy above is specific to exchange-coupled, collective-readout protocols. Each gap has a distinct physical origin: the $3\,\mathrm{dB}$ gap between the Heisenberg limit and $N^2/2$ is a property of the OAT probe state; the additional $1.34\,\mathrm{dB}$ from $N^2/2$ to $N^2/e$ is the geometric cost of the collective $\langle J_y\rangle$ readout. The total $6\,\mathrm{dB}$ gap for the butterfly echo decomposes as $4.77\,\mathrm{dB}$ from distributing QFI uniformly over all three axes (isotropy cost: $N^2 \to N^2/3$) plus $1.25\,\mathrm{dB}$ from the $3/4$ echo extraction efficiency of the $\langle J_z\rangle$ readout ($N^2/3 \to N^2/4$). The $1.66\,\mathrm{dB}$ protocol-level difference between SATIN and the butterfly echo reflects this decomposition: a $4.77\,\mathrm{dB}$ isotropy penalty is partially recovered by the more efficient echo extraction, with the net result being the $1.66\,\mathrm{dB}$ offset. These limits are intrinsic to the protocols and are not consequences of the central-spin architecture. The CSM introduces only finite Schrieffer--Wolff corrections that vanish in the dispersive limit.

%%=============================================================
\section{Summary and Outlook}
\label{section:conclusion}
%%=============================================================

In this work, we have demonstrated that the central spin model supports two distinct regimes of quantum-enhanced metrology within the same hardware architecture. In the dispersive regime, the central spin mediates an effective one-axis twisting interaction, enabling the SATIN echo protocol to achieve a QFI of approximately $N^2/e$ using the collective readout. The SATIN performance degrades measurably with Schrieffer--Wolff errors at large $N$, while the butterfly echo ceiling tracks the Haar limit even where the OAT mapping is imperfect, because its ceiling depends only on isotropy within the Dicke subspace. Furthermore, we provide a blueprint for realizing the butterfly echo protocol in a concrete physical Hamiltonian model. This protocol reaches a QFI of approximately $N^2/4$ for arbitrary unknown encoding axes using an essentially $N$-independent number of steps ($8$) over the values of $N$ considered in this work.

A second independent result of this work concerns the role of the mediator spin magnitude $S_{\rm cen}$. We have shown a clean separation of resources: increasing $S_{\rm cen}$ reduces the protocol timescale by a factor $\propto(2S_{\rm cen})^{-1}$ without altering the metrological ceiling, the optimal squeezing angle, or the circuit depth required to reach isotropy. Speed and randomness are therefore independent resources, and the mediator spin controls only the former. The proposed architecture is relevant to cavity-QED implementations with dispersively coupled atomic ensembles~\cite{Colombo_2022}, as well as NMR quantum processors~\cite{asthana2026efficientmanybodyshadowmetrology}.

Several natural directions remain for future investigation. An important next step is to incorporate realistic decoherence mechanisms in order to assess the performance of both protocols under experimentally relevant coherence times \cite{PhysRevA.87.022337}. Furthermore, extending the analysis to inhomogeneous couplings, such as the dipolar interactions encountered in NV-center platforms, would clarify the robustness of the Dicke-subspace dynamics against spatial disorder and the extent to which available dynamical-decoupling techniques can mitigate such inhomogeneity~\cite{PhysRevB.99.174308,
PhysRevApplied.17.044028}. Finally, the butterfly echo has applications beyond quantum sensing: its random scrambling circuit efficiently prepares approximate Haar-random states on the symmetric Dicke subspace, providing a promising resource for quantum information protocols such as quantum cryptography~\cite{bringewatt2026butterflyechoprotocolaxisagnostic}. The central spin architecture developed here offers a direct route to realizing these applications.

\section*{Acknowledgements}

SC acknowledges funding under the Government of India’s National Quantum Mission grant numbered DST/FFT/NQM/QSM/2024/3. AK was partially supported by the Master's Thesis Visiting Students Program at HRI.

\appendix

%%=============================================================
\section{Effective Hamiltonian Derivation}
\label{app:eff_ham}
%%=============================================================
 
We derive the effective bath Hamiltonian via second-order
perturbation theory (Schrieffer--Wolff transformation).
Starting from the full central spin model
\begin{equation}
  H = \Omega S_z^{(c)}
    + \frac{\tilde{g}}{2}
      \!\left( S_+^{(c)} I_-^{(b)} + S_-^{(c)} I_+^{(b)} \right),
  \label{eq:H_full}
\end{equation}
we identify the unperturbed part $H_0 = \Omega S_z^{(c)}$ with
eigenenergies $E_m = \Omega m$, and treat the exchange coupling as
the perturbation
$V = \tfrac{\tilde{g}}{2}(S_+^{(c)} I_-^{(b)} + S_-^{(c)} I_+^{(b)})$.
 
The central spin is initialized in its lowest-energy eigenstate
$\ket{-S_{\rm cen}}$, giving $E_0 = -\Omega S_{\rm cen}$.
Acting $V$ on the initial state $\ket{-S_{\rm cen}}\otimes\ket{\Psi_{\rm bath}}$:
\begin{align}
  V\bigl(\ket{-S_{\rm cen}}\otimes\ket{\Psi_{\rm bath}}\bigr)
  &= \frac{\tilde{g}}{2}
     \sqrt{2S_{\rm cen}}\,
     \ket{-S_{\rm cen}+1}\otimes I_-^{(b)}\ket{\Psi_{\rm bath}},
  \label{eq:V_action}
\end{align}
where we used $S_+^{(c)}\ket{-S_{\rm cen}} = \sqrt{2S_{\rm cen}}\,
\ket{-S_{\rm cen}+1}$ and $S_-^{(c)}\ket{-S_{\rm cen}} = 0$.
Only the single virtual level $m' = -S_{\rm cen}+1$ contributes.
Its energy denominator is
\begin{equation}
  E_0 - E_{m'} = -\Omega S_{\rm cen} - \Omega(-S_{\rm cen}+1) = -\Omega.
  \label{eq:energy_denom}
\end{equation}

\begin{table*}[t]
    \centering
    \caption{Comparison of peak metrological gain for the ideal
effective collective-twisting model and the CSM at varying
bath sizes $N$ and detunings $\Omega$
($\tilde{g}=2$, $S_{\rm cen}=0.5$). All values are in dB.
The ``Gap from ideal'' column quantifies the additional
dynamical SW penalty beyond the intrinsic finite-$N$ correction
of the collective $\langle J_y\rangle$ echo readout.}
    \renewcommand{\arraystretch}{1.2}
    \begin{tabular}{cc | c | cc | cc}
        \hline\hline
        $N$ & HL & Ideal SATIN echo ($\langle J_y\rangle$) & CS ($\Omega=1000$) & Gap from Ideal &
        CS ($\Omega=2000$) & Gap from Ideal \\
        \hline
        100  & 20.00 & 15.64 & 15.50 & 0.14 & 15.54 & 0.10 \\
        200  & 23.01 & 18.66 & 18.43 & 0.23 & 18.57 & 0.09 \\
        300  & 24.77 & 20.42 & 20.01 & 0.41 & 20.29 & 0.13 \\
        400  & 26.02 & 21.67 & 20.99 & 0.68 & 21.48 & 0.19 \\
        500  & 26.99 & 22.64 & 21.81 & 0.83 & 22.38 & 0.26 \\
        800  & 29.03 & 24.69 & 22.51 & 2.18 & 24.02 & 0.67 \\
        1000 & 30.00 & 25.65 & 23.23 & 2.42 & 24.88 & 0.77 \\
        \hline\hline
    \end{tabular}
    \label{tab:gain_comparison}
\end{table*}
 
Projecting onto the central-spin ground state, the bath operator
obtained from second-order perturbation theory is
\begin{align}
  H_{\rm eff}^{(b)}
  &= \sum_{m' \neq -S_{\rm cen}}
     \frac{
       \langle -S_{\rm cen}\rvert V \lvert m' \rangle\,
       \langle m' \rvert V \lvert -S_{\rm cen} \rangle
     }{E_0 - E_{m'}}
  \notag \\
  &= \frac{(\tilde{g}/2)^2 \cdot 2S_{\rm cen}}{-\Omega}
     \; I_+^{(b)} I_-^{(b)}
  = -\frac{\tilde{g}^2 S_{\rm cen}}{2\Omega}\,I_+^{(b)} I_-^{(b)}.
  \label{eq:H_eff_raw}
\end{align}
Here $\langle m'|V|-S_{\rm cen}\rangle$ denotes the central-spin
matrix element, whose bath-operator part is $I_-^{(b)}$ for $m'
= -S_{\rm cen}+1$.
 
Using the angular-momentum identity $I_+^{(b)}I_-^{(b)} =
\mathbf{I}^2 - (J_z)^2 + J_z$ and substituting into
Eq.~(\ref{eq:H_eff_raw}):
\begin{equation}
  H_{\rm eff}^{(b)} = -\frac{\tilde{g}^2 S_{\rm cen}}{2\Omega}
  \!\left[\mathbf{I}^2 - (J_z)^2 + J_z\right].
\end{equation}
Within the symmetric Dicke subspace, $\mathbf{I}^2 = \frac{N}{2}
\!\left(\frac{N}{2}+1\right)\mathbf{1}$ is a constant, and the
linear term $J_z$ produces only a global precession that
does not generate entanglement.  Dropping both, the
$-(- (J_z)^2)$ term changes sign, leaving the effective
One-Axis Twisting (OAT) Hamiltonian~\cite{PhysRevA.47.5138}:
\begin{equation}
    H_{\rm eff}^{(b)} = +\chi_{\rm eff}\!\left(J_z\right)^{\!2},
    \qquad
    \chi_{\rm eff} = \frac{\tilde{g}^2 S_{\rm cen}}{2\Omega}.
  \label{eq:OAT}
\end{equation}
With the fixed physical coupling $\tilde{g}=2$ used throughout
this work, this reduces to $\chi_{\rm eff} = 2S_{\rm cen}/\Omega$.
The positive sign confirms that the bath spins undergo genuine
one-axis twisting.  The effective twisting rate is proportional to
$S_{\rm cen}$: a larger mediator spin accelerates entanglement
generation without affecting the ceiling itself,
since the peak gain depends on $N$ alone.

The exact peak gain of a pure SATIN echo at finite $N$ is obtained by maximizing
the signal-to-noise ratio $\mathcal{S}^2/(N/4)$ where
$\mathcal{S}(t) = (N/2)(N-1)\sin(\chi t)\cos^{N-2}(\chi t)$.
Setting $d\mathcal{S}/dt = 0$ gives $\tan^2(\chi t_{\rm opt}) = 1/(N-2)$, and
substituting back yields:
\begin{equation}
\label{eq:echo_finite_N}
    \mathcal{G}_{\rm echo}^{\rm SATIN}(N)
    = 10\log_{10}\!\left[(N-1)
      \left(\frac{N-2}{N-1}\right)^{N-2}\right].
\end{equation}
This differs from the asymptotic $10\log_{10}(N/2)$ by approximately
$-1.34\,\mathrm{dB}$ for all $N \geq 100$, a correction that is intrinsic to the
echo readout and cannot be closed by tuning any Hamiltonian parameter.
The additional SW penalty listed in Table~\ref{tab:gain_comparison} is entirely
recoverable by increasing $\Omega$; reaching $<0.5\,\mathrm{dB}$ SW error at
$N = 1000$ would require $\Omega \gtrsim 5000$.

\begin{figure*}
    \centering
    \includegraphics[width=0.7\textwidth]{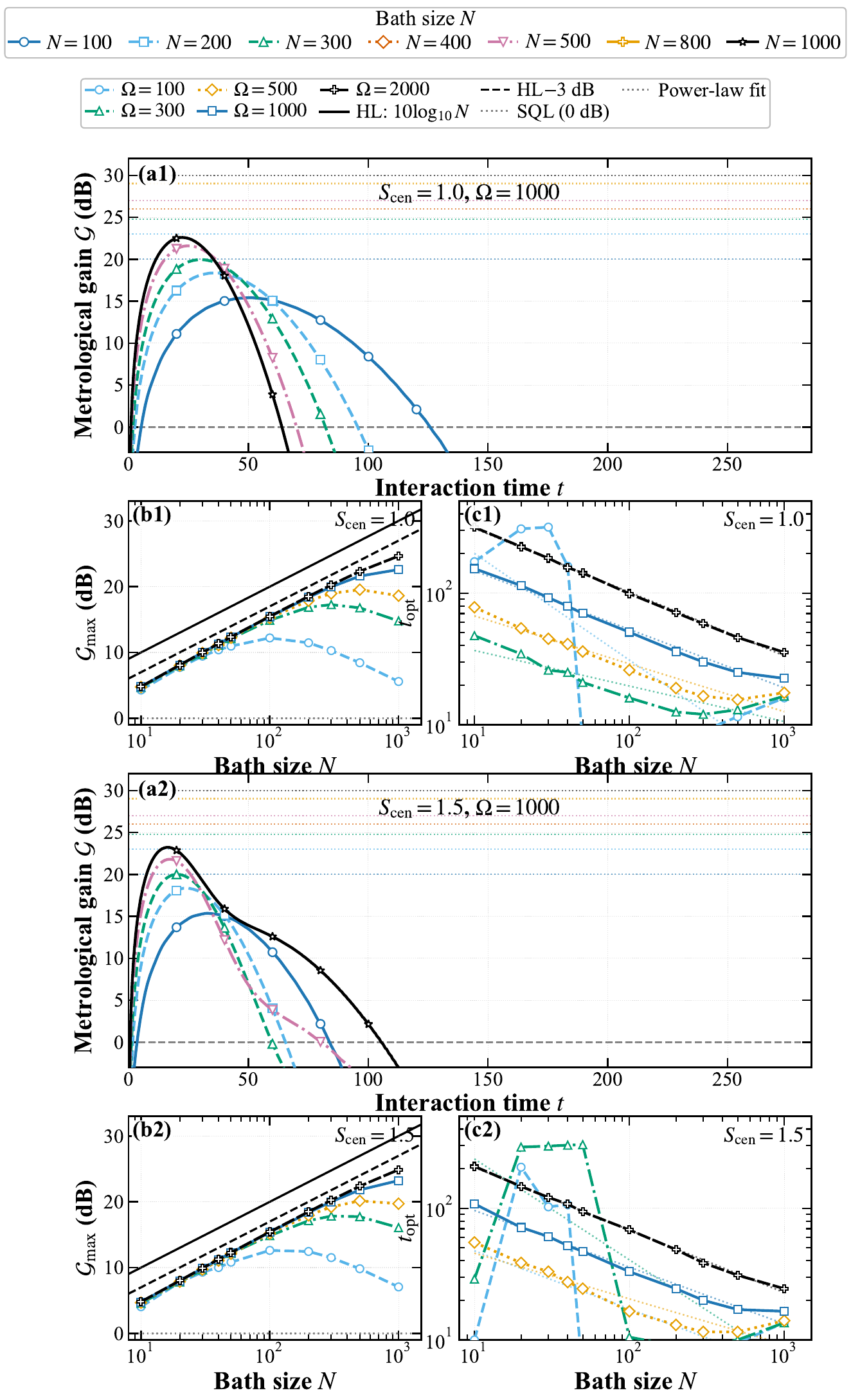}
    \caption{%
        Standard echo gain dynamics and scaling for
        $S_{\rm cen}=1.0$ (top row) and $S_{\rm cen}=1.5$ (bottom row),
        with $\tilde{g}=2$ and $\Omega=1000$.
        Panels $(a1)$ and $(a2)$ show the gain dynamics for varying bath size
        $N$. Panels $(b1)$ and $(b2)$ show the peak gain scaling with $N$,
        while panels $(c1)$ and $(c2)$ show the corresponding optimal
        interaction time $t_{\rm opt}$.
        The peak gain is independent of $S_{\rm cen}$, while increasing
        $S_{\rm cen}$ reduces the optimal time by the corresponding factors,
        consistent with Eq.~\eqref{eq:topt_CSM}.
    }
    \label{fig:standard_echo_comparison}
\end{figure*}

%%=============================================================
\section{Standard Echo Protocol with CSM}
\label{app:scen}
%%=============================================================

\subsection{Phase Sensitivity}
\label{app:deltaphi}

In this subsection, we derive the analytical phase sensitivity predicted by the effective OAT model, which describes the full CSM dynamics in the dispersive regime. Defining the SQL as $(\Delta\phi)^2_{\rm SQL} = 1/N$, the metrological gain is (Eq.~\eqref{eq:gain_dB} of the main text):
\begin{equation}
  \text{Gain (dB)} = 10\log_{10}\!\left(\frac{1}{N(\Delta\phi)^2}\right).
\end{equation}
 
At the echo readout point ($\phi = 0$), the time-reversal maps
the bath state back to the coherent spin state (CSS)
$\ket{+x}^{\otimes N}$.  The uncertainty in the spin
measurement originates from the intrinsic CSS projection
noise $\Delta S_{\rm quant} = \sqrt{N/4}$, leading to
\begin{equation}
  \Delta\phi =
  \frac{\sqrt{(\Delta S_{\rm quant})^2}}
       {\left|\partial_\phi \langle J_y^\phi\rangle\big|_{\phi=0}\right|}
  = \frac{\sqrt{N/4 }}
         {\mathcal{S}(t)},
  \label{eq:deltaphi}
\end{equation}
where $\mathcal{S}(t) \equiv |\partial_\phi \langle J_y^\phi \rangle|_{\phi=0}|$ is the signal slope evaluated below.
 
For the effective OAT calculation, the bath echo final state is
\[
\ket{\psi_{\rm out}}
=
U_{\rm OAT}^\dagger e^{-i\phi J_y}
U_{\rm OAT}\ket{+x}^{\otimes N},
\qquad
U_{\rm OAT}=e^{-i\chi_{\rm eff}tJ_z^2}.
\]
Expanding the rotation operator to first order in $\phi$ and using
$\langle +x|J_y|+x\rangle=0$:
\begin{equation}
  \left.
  \frac{\partial\langle J_y^\phi\rangle}{\partial\phi}
  \right|_{\phi=0}
  =
  i\,\langle +x \rvert
  \bigl[U_{\rm OAT}^\dagger J_y U_{\rm OAT},J_y\bigr]
  \lvert +x\rangle.
  \label{eq:slope_commutator}
\end{equation}
For the OAT model $H = \chi J_z^2$ with $J = N/2$, this
commutator evaluates analytically~\cite{PhysRevLett.116.053601}
to yield the signal slope
\begin{equation}
  \mathcal{S}(t) = J(2J-1)\sin(\chi t)\cos^{2J-2}(\chi t).
  \label{eq:slope_OAT}
\end{equation}
In the dispersive limit of the central-spin model, substituting
$J=N/2$ and $\chi\to\chi_{\rm eff}=2S_{\rm cen}/\Omega$
for $\tilde{g}=2$ gives the analytical OAT prediction:
\begin{equation}
  \mathcal{S}(t) = \frac{N}{2}(N-1)
    \sin\!\left(\frac{2S_{\rm cen}\,t}{\Omega}\right)
    \cos^{N-2}\!\!\left(\frac{2S_{\rm cen}\,t}{\Omega}\right).
  \label{eq:slope_CSM}
\end{equation}
 
Substituting Eq.~(\ref{eq:slope_CSM}) into Eq.~(\ref{eq:deltaphi})
gives the phase sensitivity predicted by the effective OAT model
in the dispersive regime:
 
\begin{equation}
  \Delta\phi =
  \dfrac{\displaystyle\sqrt{\frac{N}{4}}}
  {\dfrac{N}{2}(N-1)
   \sin\!\!\left(\dfrac{2S_{\rm cen}\,t}{\Omega}\right)
   \cos^{N-2}\!\!\!\left(\dfrac{2S_{\rm cen}\,t}{\Omega}\right)}.
  \label{eq:deltaphi_full}
\end{equation}
In the large-$N$ limit, this approaches a maximum Fisher information $N^2/e$ with the collective $\langle J_y\rangle$ readout, corresponding to a metrological gain $10\log_{10}N-4.34\,\mathrm{dB}$ below the Heisenberg limit. We note that this difference from the Heisenberg limit arises due to two distinct physical mechanisms. The first contribution is the intrinsic penalty associated with the
collective readout, while the second arises from finite
Schrieffer--Wolff corrections in the CSM. As discussed in
Sec.~\ref{subsec:results_echo}, these corrections reduce the signal
slope rather than the projection-noise floor of the echo output.

\subsection{Optimal Protocol Timescale}
\label{app:topt}
 
We determine the squeezing time $t_{\rm opt}$ that maximises
the echo signal slope $\mathcal{S}(t)$.
 
Substituting $x \equiv \chi t$ into Eq.~(\ref{eq:slope_OAT}),
we require $d\mathcal{S}/dx = 0$:
\begin{align}
  \frac{d}{dx}\!\left[\sin(x)\cos^{2J-2}(x)\right]
  & = 0.
  \notag
\end{align}
\begin{equation}
  \cos^2(x) = (2J-2)\sin^2(x)
  \quad\Longrightarrow\quad
  \tan^2(x_{\rm opt}) = \frac{1}{2J-2}.
  \label{eq:topt_condition}
\end{equation}
For large $J = N/2 \gg 1$, Eq.~(\ref{eq:topt_condition}) gives
$x_{\rm opt} \approx 1/\sqrt{2J}$, so the optimal dimensionless
twisting time is
\begin{equation}
  \chi t_{\rm opt} \approx \frac{1}{\sqrt{2J}} \propto N^{-1/2}.
  \label{eq:chi_topt}
\end{equation}
This gives the physical SATIN echo optimal time
\begin{equation}
  t_{\rm opt}^{\rm SATIN} \approx \frac{1}{\chi\sqrt{2J}}
  = \frac{1}{\chi\sqrt{N}}.
  \label{eq:topt_SATIN}
\end{equation}
Note that Eq.~(\ref{eq:chi_topt}) correctly decreases with $N$:
larger ensembles reach their optimal entanglement faster on the
dimensionless clock $\chi t$.  This contrasts with the direct OAT
no-echo optimal time $\chi t \sim N^{-2/3}$ (Kitagawa-Ueda~\cite{PhysRevA.47.5138}),
from which Eq.~(\ref{eq:chi_topt}) shows that the echo protocol
operates in the over-squeezed regime, converting
otherwise inaccessible QFI into measurable signal.
 
Substituting $\chi\to\chi_{\rm eff} = 2S_{\rm cen}/\Omega$
and $J = N/2$ into Eq.~(\ref{eq:topt_SATIN}):
\begin{equation}
    t_{\rm opt}^{\rm CSM}
    \approx \frac{1}{\chi_{\rm eff}\sqrt{N}}
    = \frac{\Omega}{2S_{\rm cen}\sqrt{N}}
    \propto \frac{\Omega}{S_{\rm cen}\,N^{1/2}}.
  \label{eq:topt_CSM}
\end{equation}
Equation~(\ref{eq:topt_CSM}) encodes the two independent
experimental control knobs: the detuning $\Omega$ sets the
macroscopic timescale linearly, while $S_{\rm cen}$ provides a
multiplicative fine-tuning through the factor $(2S_{\rm cen})^{-1}$,
so that a spin-$3/2$ mediator runs three times faster than a
spin-$1/2$ qubit at the same $\Omega$.
Since the peak gain is $S_{\rm cen}$-independent,
the ancilla size functions purely as a speed knob without any
trade-off in metrological performance.

%%=============================================================
\section{Butterfly Echo with CSM Implementation}
\label{app:butterfly}
%%=============================================================

\begin{figure}[htbp]
    \centering
    \includegraphics[width=0.9\columnwidth]{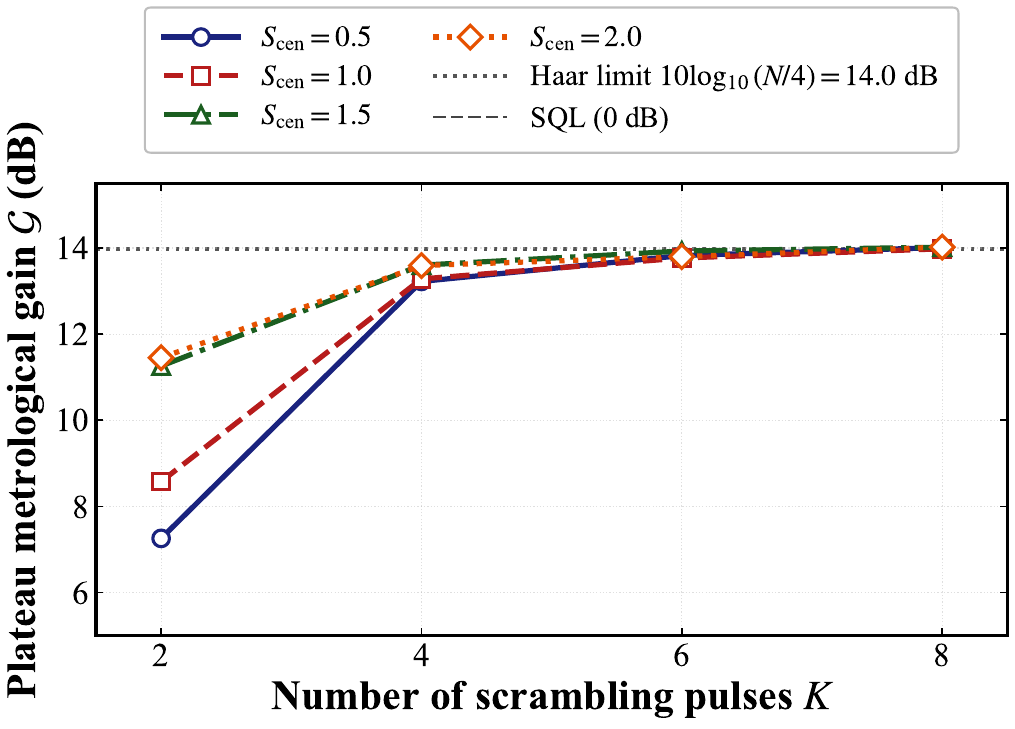}
    \caption{%
        Plateau gain as a function of the number of random scrambling pulses $K$,
        for ancilla spin magnitudes $S_{\rm cen} \in \{0.5, 1.0, 1.5, 2.0\}$.
        All curves converge to the same Haar limit ($\approx 14.0\,\mathrm{dB}$,
        black dotted line) by $K = 8$.
        Parameters: $N = 100$, $\Omega = 1000$, $\tilde{g} = 2$, averaged over
        50 random circuit realisations.    }
    \label{fig:plateau_vs_K}
\end{figure}

\begin{figure*}[htbp]
    \centering
    \begin{subfigure}{0.45\textwidth}
        \centering
        \includegraphics[width=\linewidth]{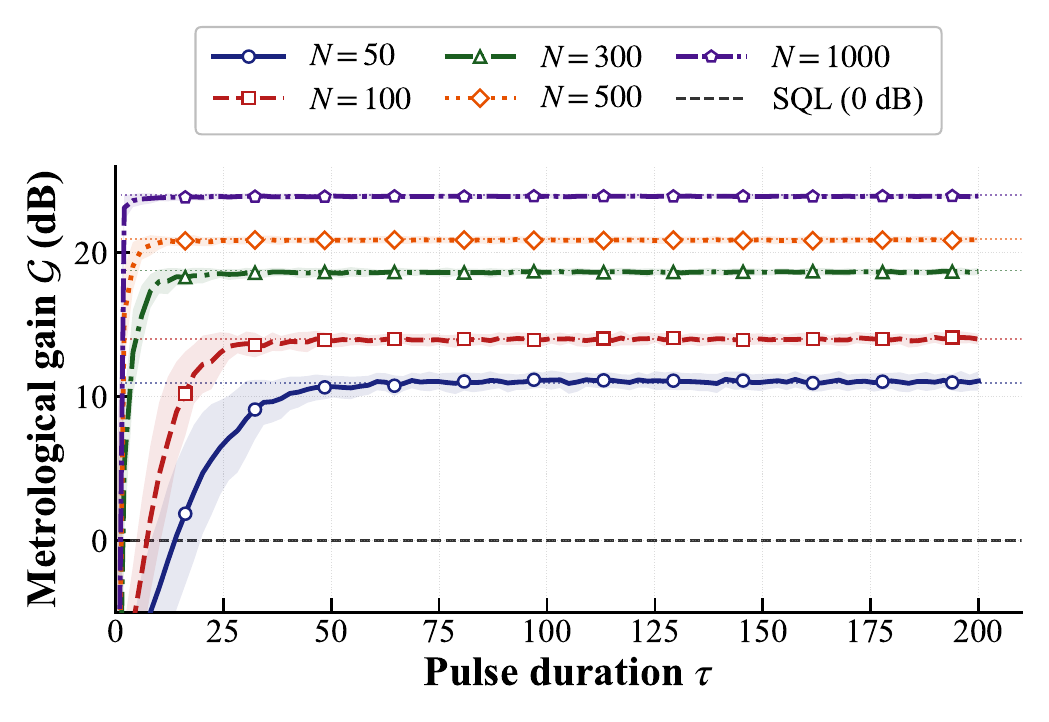}
        \caption{$S_{\rm cen}=1.0$: Varying $N$ ($\Omega=1000$)}
        \label{fig:bfly_sc1_gvt_N}
    \end{subfigure}\hfill
    \begin{subfigure}{0.45\textwidth}
        \centering
        \includegraphics[width=\linewidth]{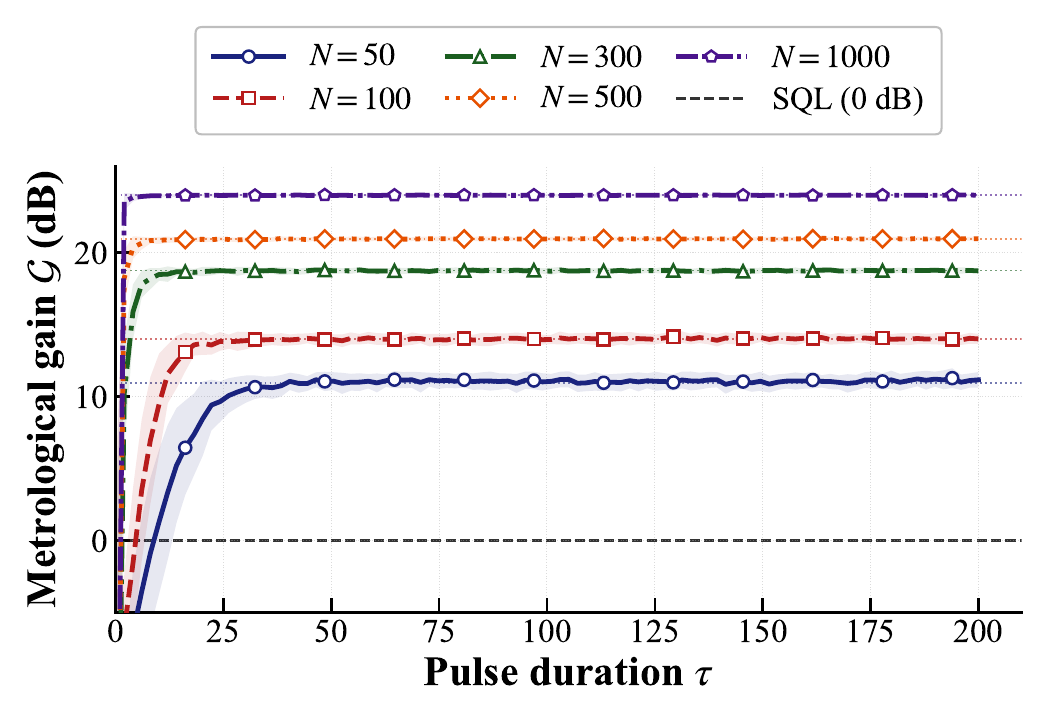}
        \caption{$S_{\rm cen}=1.5$: Varying $N$ ($\Omega=1000$)}
        \label{fig:bfly_sc15_gvt_N}
    \end{subfigure}
    
    \vspace{0.4cm}
    
    \begin{subfigure}{0.45\textwidth}
        \centering
        \includegraphics[width=\linewidth]{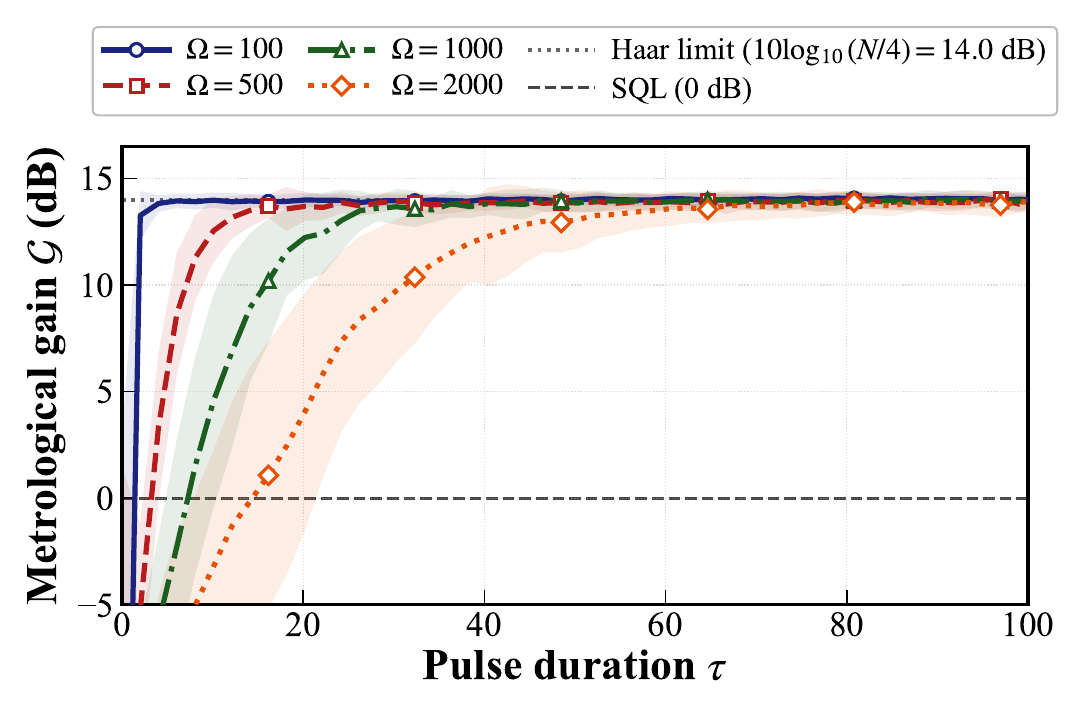}
        \caption{$S_{\rm cen}=1.0$: Varying $\Omega$ ($N=100$)}
        \label{fig:bfly_sc1_gvt_om}
    \end{subfigure}\hfill
    \begin{subfigure}{0.45\textwidth}
        \centering
        \includegraphics[width=\linewidth]{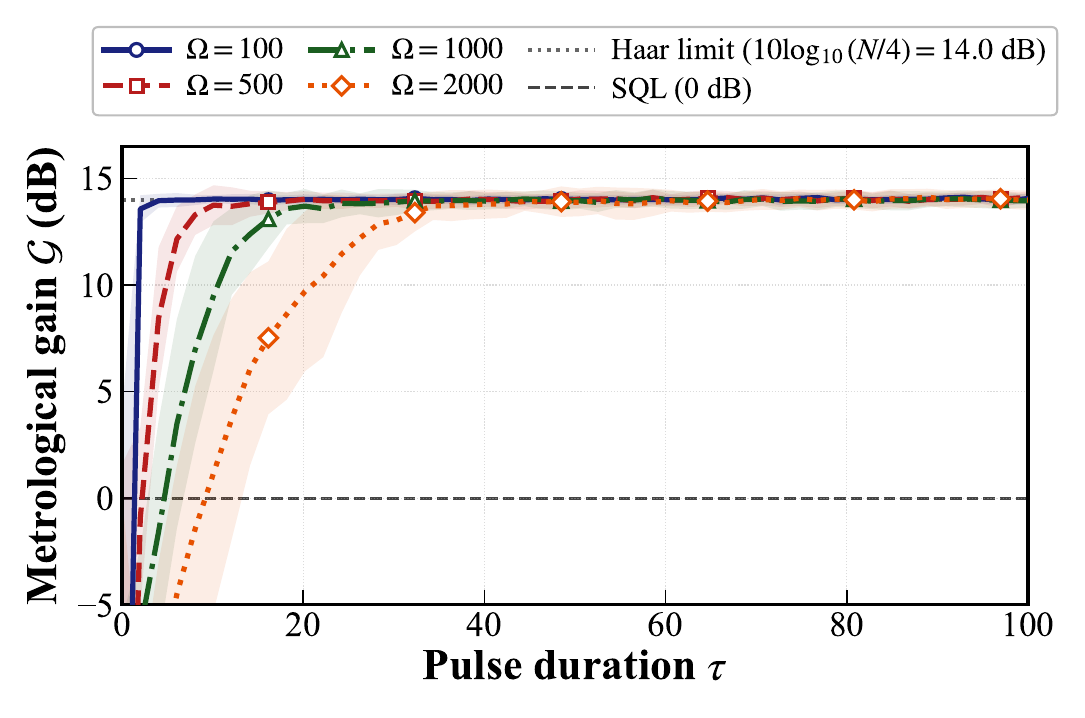}
        \caption{$S_{\rm cen}=1.5$: Varying $\Omega$ ($N=100$)}
        \label{fig:bfly_sc15_gvt_om}
    \end{subfigure}

    \vspace{0.4cm}
    
    \begin{subfigure}{0.45\textwidth}
        \centering
        \includegraphics[width=\linewidth]{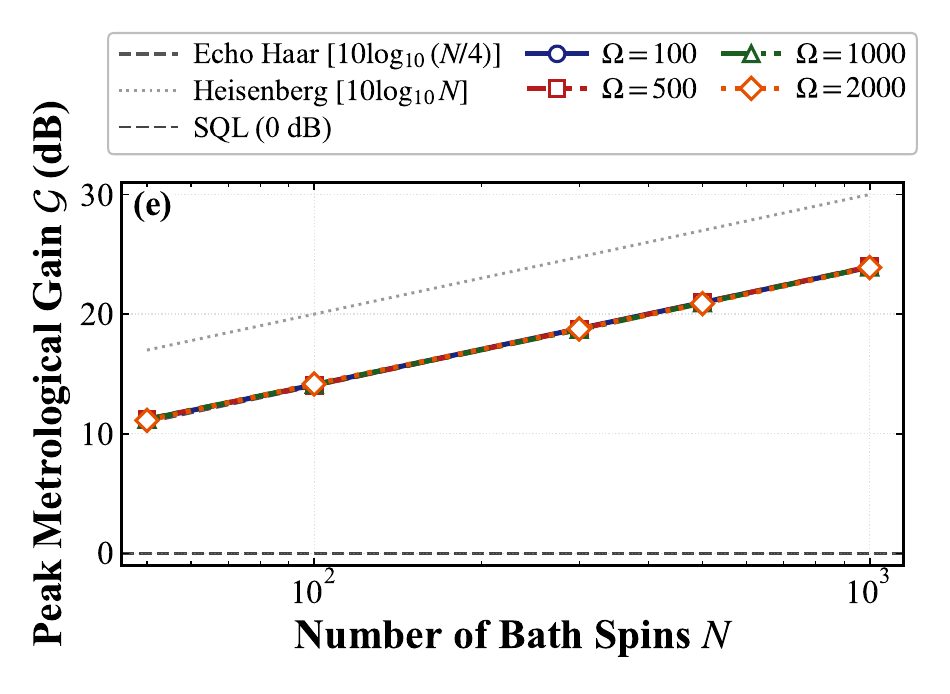}
        \caption{$S_{\rm cen}=1.0$: Peak gain scaling vs $N$}
        \label{fig:bfly_sc1_scaling}
    \end{subfigure}\hfill
    \begin{subfigure}{0.45\textwidth}
        \centering
        \includegraphics[width=\linewidth]{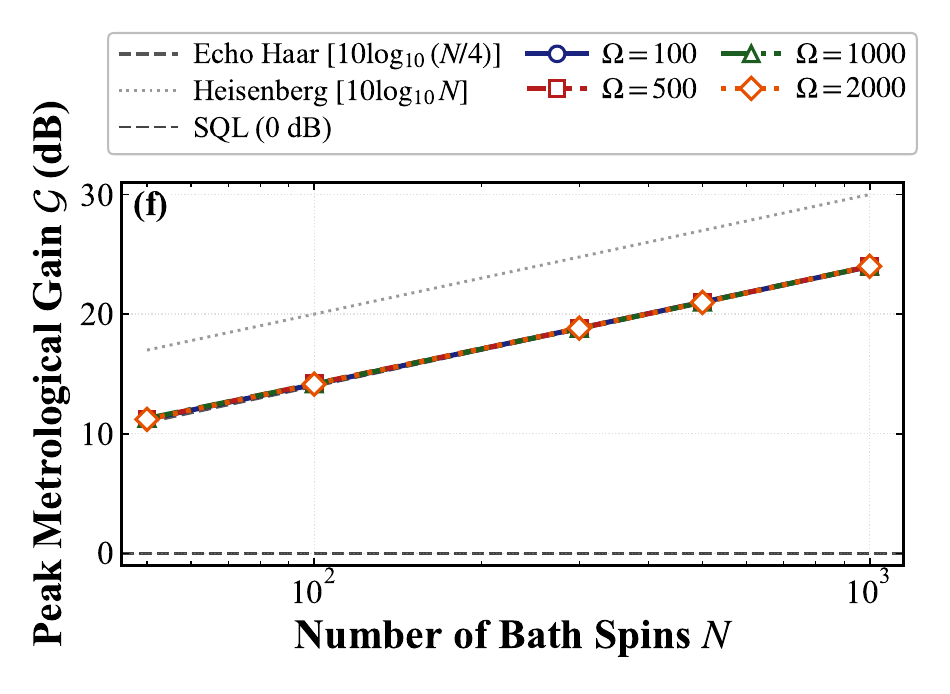}
        \caption{$S_{\rm cen}=1.5$: Peak gain scaling vs $N$}
        \label{fig:bfly_sc15_scaling}
    \end{subfigure}
    
    \caption{%
        Extended butterfly echo results for $S_{\rm cen}\in\{1.0,1.5\}$,
        mirroring Fig.~\ref{fig:butterfly_dynamics} for $S_{\rm cen}=0.5$.
        \textbf{(a,\,b)}~Gain dynamics at $\Omega=1000$ for varying $N$:
        rise-to-plateau profiles saturate at the Haar limit
        $10\log_{10}(N/4)$ (Eq.~\eqref{eq:haar_limit}), with $\tau_{\rm opt}$
        reduced by factors of $2$ and $3$ relative to $S_{\rm cen}=0.5$.
        \textbf{(c,\,d)}~Gain dynamics at $N=100$ for varying $\Omega$,
        illustrating the onset of the dispersive regime; curves at large $\Omega$
        approach the Haar limit within the dispersive error (Table~\ref{tab:epsilon}).
        \textbf{(e,\,f)}~Peak gain scaling vs $N$: all curves track the Haar limit
        to within $0.3\,\mathrm{dB}$, confirming $S_{\rm cen}$-independence of the
        metrological ceiling.
    }
    \label{fig:appendix_butterfly_higher_spins}
\end{figure*}
In this appendix, we first summarize the principal theoretical results of the butterfly echo protocol for completeness. The detailed derivations are given in the Supplemental Material of Ref.~\cite{bringewatt2026butterflyechoprotocolaxisagnostic}.
We then present the new result specific to this work: a quantification of the
dispersive approximation error introduced by the central spin implementation.

\textit{Unknown-axis channel} :
When the encoding axis $\hat{n}$ is unknown, the encoding is described by the
mixed-unitary depolarizing
channel~\cite{bringewatt2026butterflyechoprotocolaxisagnostic}
(Eq.~\eqref{eq:depolarising_channel}),
whose QFI at leading order in $\theta$ is (Eq.~\eqref{eq:QFI_depolarising}):
\begin{equation}
  F(\theta) = \frac{4}{3}
    \sum_{j\in\{x,y,z\}}\mathrm{Var}(J_j).
\end{equation}
Heisenberg-limited axis-agnostic sensing therefore requires the probe state
covariance matrix to be proportional to the identity.

\textit{Signal shape} :
For a Haar-random probe state the ensemble-averaged output polarization follows a
$\mathrm{sinc}^2$ profile (Eq.~\eqref{eq:butterfly_signal})
\cite{bringewatt2026butterflyechoprotocolaxisagnostic}:
\begin{equation}
  \frac{\overline{\langle J_z\rangle}}{I}
  \;\xrightarrow{N\to\infty}\;
  \mathrm{sinc}^2(I\theta),
  \qquad I = N/2,
\end{equation}
with metrological bandwidth $\theta_{\rm bw} = 2\pi/N$.
The quadratic dependence on $\theta$ near $\theta=0$ is a direct consequence of the
$\hat{n} \to -\hat{n}$ symmetry of the depolarizing channel (discussed in
Sec.~\ref{subsec:butterfly}).

\textit{Echo Haar limit} :
Averaging $F(\theta)$ over Haar-random probe states in the symmetric subspace
$\mathcal{S}_N$ and accounting for the $z$-axis echo readout
efficiency~\cite{bringewatt2026butterflyechoprotocolaxisagnostic}:
\begin{equation}
  \mathcal{G}_{\rm Haar}^{\rm echo}
  = 10\log_{10}\!\left(\frac{N}{4}\right),
\end{equation}
exactly $6\,\mathrm{dB}$ below the Heisenberg limit and
$3\,\mathrm{dB}$ below the squeezed-state parity-readout
benchmark.
Both protocols share the same $10\,\mathrm{dB/decade}$ Heisenberg slope. The $6\,\mathrm{dB}$ offset decomposes into
$4.77\,\mathrm{dB}$ from distributing QFI uniformly over the three
axes ($N^2\to N^2/3$), followed by $1.25\,\mathrm{dB}$ from the
$3/4$ echo-extraction efficiency ($N^2/3\to N^2/4$).

\textit{Circuit depth} :
A random OAT circuit of depth $K$ converges to an approximate Haar measure on
$\mathcal{S}_N$ in $K_{\rm sat} = \mathcal{O}(1)$ steps, because
the effective Lindbladian governing the moment evolution has a spectral gap constant in $N$~\cite{bringewatt2026butterflyechoprotocolaxisagnostic}.
In the present calculations, $K=8$ suffices to reach the Haar-limit performance within $0.3\,\mathrm{dB}$ for all system sizes studied here, consistent with that prior result.

\paragraph{Plateau vs K convergence:} Figure~\ref{fig:plateau_vs_K} demonstrates that $S_{\rm cen}$ does not reduce the number of random OAT
steps required to generate a metrological 2-design over the symmetric subspace:
the convergence to the Haar plateau is determined entirely by $K$, the number of independently randomised axes sampled.
The ancilla spin therefore acts as a speed knob within each pulse but has no bearing on the combinatorial depth needed to achieve isotropy.

Finally, we note that the results discussed above assume ideal OAT steps.  In this work, each step is instead implemented by conjugating the physical CSM evolution with a random bath rotation (Eq.~\eqref{eq:butterfly_CS_step}).  The SW mapping (Appendix~\ref{app:eff_ham})
gives:
\begin{equation}
  U_j^{\rm CSM}(\hat{m}_j)
  = e^{-i\chi_{\rm eff}\tau(J_{\hat{m}_j})^2}
    \Bigl[1 + \mathcal{O}(\epsilon)\Bigr],
  \qquad
\
\epsilon =
\left(
\frac{2\tilde{g}\sqrt{N}}{\Omega}
\right)^2,
  \label{eq:CSM_step_error}
\end{equation}
where $\epsilon$ is the fractional per-step deviation from an ideal OAT twist.
Table~\ref{tab:epsilon} lists $\epsilon$ for the parameter combinations used in
this work.

\begin{table}[h]
\centering
\caption{Per-step dispersive error $\epsilon =
4(\tilde{g}\sqrt{N}/\Omega)^2$ for $\tilde{g}=2$, consistent with the tabulated entries.
Accumulated error after $K=8$ steps is $8\epsilon$.}
\label{tab:epsilon}
\begin{tabular}{ccccc}
\hline\hline
$N$ & $\Omega=100$ & $\Omega=500$ & $\Omega=1000$ & $\Omega=2000$ \\
\hline
100  & 0.160 & 0.006 & 0.002 & 0.001 \\
300  & 0.480 & 0.019 & 0.005 & 0.001 \\
500  & 0.800 & 0.032 & 0.008 & 0.002 \\
1000 & 1.600 & 0.064 & 0.016 & 0.004 \\
\hline\hline
\end{tabular}
\end{table}

For $\Omega \geq 500$ the accumulated error $8\epsilon \leq 0.51$
for all $N \leq 1000$. From Fig.~\ref{fig:butterfly_dynamics}(b), the observed gain deficit remains within $0.3\,\mathrm{dB}$ of the Haar limit for $\Omega \geq 500$.
At $\Omega = 100$ the error grows large at $N \gtrsim 500$, yet
the plateau gain remains close to the Haar limit because the
non-OAT corrections introduced by the SW error are themselves
random and non-commuting, contributing additional scrambling
rather than coherently suppressing it.

\bibliography{ref}

\end{document}